\documentclass[a4paper]{jpconf}
\usepackage{graphicx}
\begin{document}
\newcommand{\nnprime}{n,n$^\prime \gamma$}
\newcommand{\ntwon}{n,2n$\gamma$}
\newcommand{\nthreen}{n,3n$\gamma$}
\newcommand{\nxn}{n,xn$\gamma$}
\newcommand{\nx}{n,x$\gamma$}

\newcommand{\natpb}{$^{\textrm{nat}}$Pb}
\newcommand{\natge}{$^{\textrm{nat}}$Ge}
\newcommand{\enrge}{$^{\textrm{enr}}$Ge}
\newcommand{\eigpb}{$^{208}$Pb}
\newcommand{\sevpb}{$^{207}$Pb}
\newcommand{\sixpb}{$^{206}$Pb}
\newcommand{\fivpb}{$^{205}$Pb}
\newcommand{\foupb}{$^{204}$Pb}
\newcommand{\nonubb}  {$0 \nu \beta \beta$}
\newcommand{\twonubb} {$2 \nu \beta \beta$}
\newcommand{\gam}{$\gamma$}
\def\nuc#1#2{${}^{#1}$#2}
\def\mee{$\langle m_{\beta\beta} \rangle$}
\def\mnu{$\langle m_{\nu} \rangle$}
\def\ml{$m_{lightest}$}
\def\gnu{$\langle g_{\nu,\chi}\rangle$}
\def\mmod{$\| \langle m_{\beta\beta} \rangle \|$}
\def\mb{$\langle m_{\beta} \rangle$}
\def\BBz{$0 \nu \beta \beta$}
\def\BBm{$\beta\beta(0\nu,\chi)$}
\def\BBt{$2 \nu \beta \beta$}
\def\nonubb{$0 \nu \beta \beta$}
\def\twonubb{$2 \nu \beta \beta$}
\def\BB{$\beta\beta$}
\def\Gz{$G^{0\nu}$}
\def\Mz{$M_{0\nu}$}
\def\Mt{$M_{2\nu}$}
\def\MzG{$M^{GT}_{0\nu}$}           
\def\MzF{$M^{F}_{0\nu}$}                
\def\MtG{$M^{GT}_{2\nu}$}           
\def\MtF{$M^{F}_{2\nu}$}                
\def\Tz{$T^{0\nu}_{1/2}$}
\def\Tt{$T^{2\nu}_{1/2}$}
\def\Tc{$T^{0\nu\,\chi}_{1/2}$}
\def\Rz{$\Gamma_{0\nu}$}            
\def\Rt{$\Gamma_{2\nu}$}            
\def\ms{$\delta m_{\rm sol}^{2}$}
\def\ma{$\delta m_{\rm atm}^{2}$}
\def\ts{$\theta_{\rm sol}$}
\def\ta{$\theta_{\rm atm}$}
\def\tot{$\theta_{13}$}
\def\gpp{$g_{pp}$}                  
\def\qval{$Q_{\beta\beta}$}                 
\def\MJ{{\sc Majorana}}             
\def\DEM{{\sc Demonstrator}}             
\def\be{\begin{equation}}
\def\ee{\end{equation}}
\def\cpRty{counts/ROI/t-y}
\def\onecpRty{1~count/ROI/t-y}
\def\fourcpRty{4~counts/ROI/t-y}
\def\ppc{P-PC}                          
\def\nsc{N-SC}           
\title{Double Beta Decay}

\author{Steven R. Elliott}

\address{Physics Division, Los Alamos National Laboratory, Los Alamos, NM 87545}

\ead{elliotts@lanl.gov}

\begin{abstract}
At least one neutrino has a mass of about 50 meV or larger. However, the absolute mass scale for the neutrino remains unknown. Furthermore, the critical question: Is the neutrino its own antiparticle? is unanswered. Studies of double beta decay offer hope for determining the absolute mass scale. In particular, zero-neutrino double beta decay (\BBz) can address the issues of lepton number conservation, the particle-antiparticle nature of the neutrino, and its mass. A summary of the recent results in \BBz, and the related technologies will be discussed in the context of the future \BBz\ program.
\end{abstract}

\section{Introduction}
Ernest Rutherford's work with $\alpha$ particles elucidated the structure of the nucleus. Scattering studies have since become
a mainstay of nuclear and particle physics research due to their ability to shed light on the structure and interactions
of the fundamental constituents of matter. This critical development in physics is why we celebrate the 100$^{th}$ anniversary of the publication describing the discovery of the atomic nucleus at this conference. In addition to scattering experiments, one can study fundamental interactions and particles through processes dependent on the exchange of virtual particles. This presentation summarizes one such process: Double Beta Decay (\BB). Two-neutrino double beta decay (\BBt) can be described as 2 neutrons simultaneously beta decaying within a nucleus emitting 2 $\beta$ particles and 2 $\nu$s. If the neutrino has certain characteristics, the alternative process of zero-neutrino double beta decay (\BBz) may occur where the neutrino is exchanged as a virtual particle between two neutrons and only $\beta$ particles are emitted in the final state. Understanding neutrino properties motivates the study of this process.

At least one neutrino has a mass of about 50 meV or larger. However, the absolute mass scale for the neutrino remains unknown. Furthermore, the critical question ``Is the neutrino its own antiparticle?" is unanswered. Studies of \BBz\ can address the issues of lepton number conservation, the particle-antiparticle nature of the neutrino, and its mass. Recent experimental results have demonstrated the increasing reach of the technologies used to search for \BBz. In addition, theoretical progress in understanding the nuclear physics involved has also been impressive. All indications are that upcoming generations of \BBz\ experiments will be sensitive to neutrino masses in the exciting range below 50 meV. 

The half-life of \BBz\ is directly related to the neutrino mass. But the half life is very long; at least greater than 10$^{25}$ years. Hence any search for the rare peak in a spectrum resulting from \BBz\ must minimize the background of other processes that may take place in a detector. Furthermore, deducing a neutrino mass value from a half-life measurement or limit requires an understanding of the transition matrix element, which is technically difficult to calculate. 

In this presentation, a summary of the the related nuclear physics of \BB\ will be discussed in the context of the future \BBz\ program. Numerous review have been written on this topic and provide great detail on this exciting science program~\cite{ell02,ell04,Avi08,Rode11,Bar11}.

\section{\BBz\ and Neutrino Mass}
The decay rate for \BBz\ can be written:

\begin{equation}
\label{eq:rate}
[T^{0\nu}_{1/2}]^{-1} = G_{0\nu} M_{0\nu}^2 \langle m_{\beta\beta} \rangle^2
\end{equation}

\noindent where $T^{0\nu}_{1/2}$ is the half-life of the decay, \Gz\ is the kinematic phase space factor, \Mz\ is the matrix element corresponding to the \BBz\ transition, and \mee\ is the effective Majorana neutrino mass. \Gz\ contains the kinematic information about the final state particles, and is exactly calculable to the precision of the input parameters (though use of different nuclear radius values in the scaling factors of \Gz\ and \Mz\ have previously introduced some confusion\cite{cow06}). 

Cosmology measures the sum of the neutrino mass eigenstates ($\Sigma$) and $\beta$ decay endpoint measurements determine a different combination ($\langle m_{\beta} \rangle$) of mass eigenvalues and neutrino mixing parameters than \BBz. The three techniques, therefore, provide complementary information on neutrino physics. The three equations are given by:

\begin{eqnarray}
\mbox{\mee} &=& \left| \sum_j m_j U_{ej}^2 \right|  \\
\langle m_{\beta} \rangle &=& \sqrt{\sum_j m_j^2 U_{ej}^2}  \\
\Sigma &=& \sum_j m_j	
\end{eqnarray}

\noindent where $m_j$ are the neutrino mass eigenstates and $U_{ej}$ are the neutrino mixing matrix elements.

\begin{figure}[h]
\includegraphics[width=25pc]{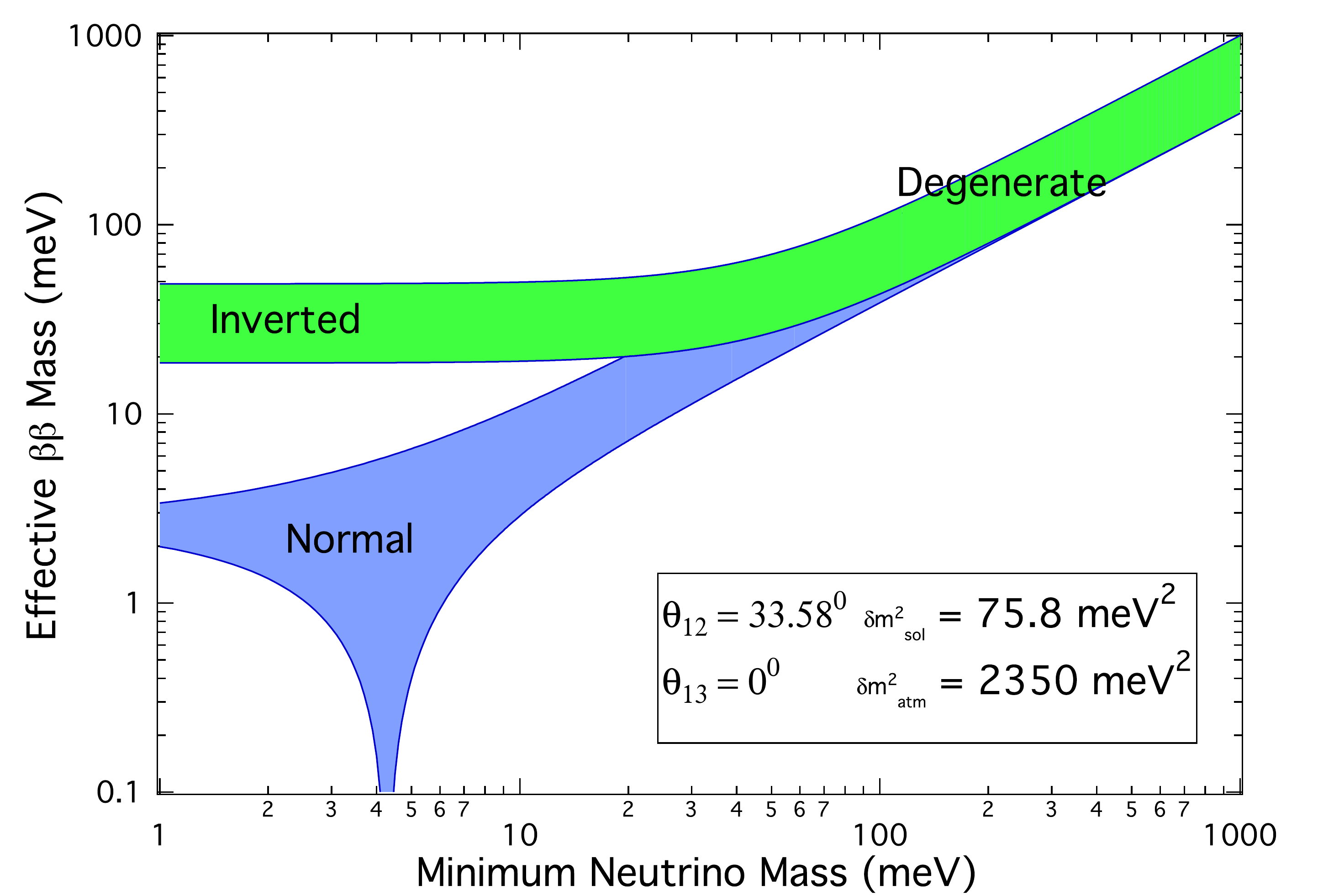}\hspace{2pc}%
\begin{minipage}[b]{10pc}\caption{\label{fig:BBmass}The effective Majorana neutrino mass as a function of the lightest neutrino mass. The neutrino oscillation parameters used for this figure come from Ref.~\cite{Fog11}}
\end{minipage}
\end{figure}

An open question in neutrino physics is whether or not the lightest neutrino mass eigenstate is the dominant component of the electron neutrino. If so, we refer to the neutrino mass spectrum as being {\em normal}. If not, we refer to it as inverted. Figure~\ref{fig:BBmass} shows the effective Majorana neutrino mass as a function of the lightest neutrino mass for these 2 possibilities. 

\section{Nuclear Physics and \BB}
The observation of \BBz\ would have profound qualitative physics conclusions.  However, the interpretation of those results quantitatively requires knowledge of \Mz.  Most nuclear matrix element calculations involve either the quasiparticle random phase approximate (QRPA) technique or the nuclear shell model (NSM).  Although the two methods have a similar starting point (a Slater determinant of independent particles), they are complementary in their treatment of correlations. QRPA uses a large number of ÒactiveÓ nucleons in a large space but with a specific type of correlation suited for collective motion.  NSM uses a small number of nucleons within a limited space but with arbitrary correlations. Three additional techniques have recently been used to look at \Mz. These include the interacting boson model (IBM-2)~\cite{Bare09}, the projected Hartree-Fock-Boboliubov (PHFB)~\cite{Chan09}, and the Energy Density Functional (EDF)~\cite{Rod10} techniques.

Recent publications~\cite{rod03, Rod06, suh05, Kor07, Fae08, Sim09a, sim09} have elucidated the causes of the historical disparity of results from QRPA calculations of \Mz.  As a result, the technique now provides a uniform result and Ref.~\cite{sim09} summarizes the values for several isotopes.  The NSM has also seen a resurgence of activity with studies of input micro-physics and its influence on \Mz ~\cite{Cau08, Cau08a} with recent values given in Ref.~\cite{Men09}.
This  progress indicates that the agreement between the various calculations is improving, but further work is needed to reduce the uncertainty to levels required for comparisons between \BBz\ results from different isotopes~\cite{Geh07,Dep07}.   

A previous workshop~\cite{zub05} was dedicated to nuclear physics studies that would support the understanding of \BB. Much work has been done since that workshop, not only on ideas presented there but others that have arisen since. Below we focus on the impact of 3 of these related nuclear physics measurements and list many additional topics.

\subsection{Atomic Masses}
The Q-values for many \BB\ transitions have not been very well known. In fact in a number of cases, experimental resolution was better than the uncertainty in the endpoint energy. Recent ion trap measurements have been providing a wealth of data to eliminate this uncertainty. (See Ref.~\cite{Eli11} for a list of references.) In the case of \nuc{130}{Te}, a new Q-value measurement~\cite{red09} resulted in an approximate 5\% \Tz\ limit change due to a shift in the endpoint energy at which the experiment searched~\cite{arn08,And11} indicating the importance of the Q-value. 

\subsection{Neutron Reactions}
Neutron reactions can result in background for \BB\ but the cross sections are often unavailable and require measurement in order to predict the impact on an experiment. For example, the \nuc{207}{Pb}(n,n'$\gamma$) reaction can produce a 3062-keV $\gamma$ ray~\cite{mei08}. This is a dangerous background for \BBz\ using \nuc{76}{Ge} because the double escape peak energy coincides with the Q-Value. Many of the cross sections for such reactions, including this example~\cite{gui09} were not well known and measurements are required. Reactions such as \nuc{76}{Ge}(n,9n)\nuc{68}{Ge} are also dangerous~\cite{Ell10} as they result from cosmic-ray induced neutrons while materials reside on the Earth's surface. The cross sections for such large $\Delta A$ transitions are small however, and many are also frequently not yet measured.

\subsection{Transfer Reactions and Pair Correlation Studies}
Proton- and neutron-removing transfer reactions have played a critical role in providing tests of the nuclear structure used for the calculation of \Mz. The difference in the nucleon configuration of the initial and final nuclei is an important input to \Mz. Reactions such as (d,p), (p,d), ($\alpha$,\nuc{3}{He}), and (\nuc{3}{He},$\alpha$) were used to study the occupation of valence neutron orbits in \nuc{76}{Ge} and \nuc{76}{Se}~\cite{Schif08}. For protons, (d,\nuc{3}{He}) reactions were studied on these isotopes~\cite{Kay09}. The correlation of pairs of neutrons were measured in this system using the (p,t) reaction~\cite{Fre07}. These data were then used to constrain calculations of \Mz\ by both the shell model~\cite{Men09a} and QRPA techniques~\cite{sim09,suh08}. As a result, the matrix element calculations of both techniques shifted such that the difference between them was reduced.

\subsection{Precise \BBt\ Data}
Accurate half-lives for \BBt~\cite{rod03} along with $\beta^{\pm}$ and electron capture decay data of the intermediate nuclei~\cite{suh05} can help determine the \gpp\ parameter used in QRPA calculations. Being able to calculate \BBt\ rates can be also considered a necessary, if not sufficient condition for calculating \BBz\ rates.

\subsection{Other Nuclear Physics Measurements for the Study of \Mz}
A variety of other proposals for studying \Mz\ have been made and we summarize them here.

{\bf Charge exchange reactions on parent \& daughter:} Charge exchange reactions, such as (p,n), (n,p), ($^3$He,t), etc., can provide data on the Gammov-Teller transition strengths of interest for \BB~\cite{eji00a}. The Fermi part of the total \BB-decay nuclear matrix element \Mz\ might be studied in charge-exchange reactions~\cite{Rod09}.

{\bf Muon capture:} The \BB\ virtual transition proceeds through levels of  the intermediate nucleus. For \BBt\ intermediate 1$^+$ states are involved, whereas for \BBz\ all J$^+$ states participate. Theoretical calculation of the relative strengths of these virtual states is very difficult. Muon capture on the final nucleus~\cite{kor02} also excites all multipoles and therefore provides  additional experimental data with which to compare calculation techniques.

{\bf Neutrino Cross Sections:} Matrix element studies could be done with neutrino beams~\cite{vol05}. By varying the average neutrino beam energy, specific multipoles of the intermediate nucleus can be excited. The strengths of both legs of the \BB\ transition could be studied by using both $\bar{\nu_e}$ and $\nu_e$.

{\bf Electromagnetic Transitions to Isobaric Analogue States:} The matrix element for the $\beta$ decay of an intermediate nucleus can be measured by observing the ($\gamma$,p) through isobaric-analog-state excitation on the daughter nucleus~\cite{eji06a}.

\section{Key Upcoming Experiments}
\label{sec:expt}
Within the next 3-5 years, a number of new experiments will begin to provide data on \BBz. All of these programs have a chance to test a recent claim of the observation of \BBz~\cite{kla06}. The validity of this claim has been debated~\cite{aal02b,fer02,har02,zde02a,Kir10} and therefore confirmation is required. Table~\ref{tab:expts} summarizes these upcoming experiments. These projects are complementary in that some emphasize isotope mass (EXO-200, KamLAND-Zen, SNO+) and others emphasize energy resolution (CUORE, GERDA, \MJ). All can potentially lead to technologies with enough isotope mass to reach sensitivity to the inverted hierarchy neutrino mass scale. It is possible that the sensitivity of these projects will result in mass limits of 100 meV or less by about 2014. Encouragingly, EXO-200 has recently claimed the observation of \BBt~\cite{Ack11}.

There are a great number of other R\&D projects that are too numerous to discuss in detail here. These include CANDLES~\cite{Kis09}, CARVEL~\cite{zde05}, COBRA~\cite{zub01}, DCBA~\cite{ish00}, LUCIFER~\cite{Giu10}, MOON~\cite{eji07}, NEXT~\cite{San09}, and SuperNEMO~\cite{Saa09}.

\begin{table}[h]
\caption{\label{tab:expts}Experiments that will provide data on \BBz\ in the near future.}
\begin{tabular}{lcccc}
\hline
Collaboration	&	Isotope			& Approx. Mass	& Run Start	&	Reference\\
\hline\hline
CUORE			&	\nuc{130}{Te}	&	200 kg		&	2013		&	\cite{Ejz09}\\
EXO-200			&	\nuc{136}{Xe}	&	200 kg		&	2011		&	\cite{Ack11}\\
GERDA			&	\nuc{76}{Ge}		&	34 kg		&	2011		&	\cite{sch05}\\
KamLAND-Zen		&	\nuc{136}{Xe}	&	400 kg		&	2011		&	\cite{Efr11}\\
\MJ\				&	\nuc{76}{Ge}		&	30 kg		&	2013		&	\cite{Aal11}\\
SNO+				&	\nuc{150}{Nd}	&	60 kg		&	2013		&	\cite{zub11}\\
\hline
\end{tabular}
\end{table}

\section{The Number of Required Experimental Results}
Although the existence of \BBz\ proves that neutrinos are massive Majorana particles~\cite{sch82}, it is not clear which lepton violating process might actually mediate the decay. Since we now know that light neutrinos do exist and would mediate \BBz\ if they are Majorana particles, that is the simplest model that could incorporate \BBz. Nevertheless, other physics might be present including: light Majorana neutrino exchange, heavy Majorana neutrino exchange, right-handed currents (RHC), and exchange mechanisms that arise from R-Parity violating supersymmetry (RPV SUSY) models.

In contrast to Eqn.~\ref{eq:rate}, the \BBz\ rate can be written more generally: 

\begin{eqnarray}\label{eq:BBRate}
[T^{0\nu}_{1/2}]^{-1} =G^{0\nu} |M_{0\nu} \eta|^{2}
\end{eqnarray}

\noindent where $\eta$ is a general lepton number violating parameter (LNVP) that was previously given by \mee.   The LNVP $\eta$ contains all of the information about lepton number violation, and has a form depending on the \BBz\ mechanism.

The LNVP takes on different forms for different \BBz\ mechanisms. In addition, \Mz\ also depends on the mechanism. The heavy-particle models represent a large class of theories that involve the exchange of high-mass ($>$1 TeV) particles.  For example, leptoquarks~\cite{Hir96b} have very similar \Mz\ to RPV SUSY~\cite{Hir96}. Left-right symmetric models can lead to right-handed current models~\cite{Doi85} or heavy neutrino exchange models~\cite{Hir96b}. Scalar bilinears~\cite{kla03c} might also mediate the decay but explicit matrix elements have not been calculated yet. For SUSY and left-right symmetric models, effective field theory~\cite{Pre03} has been used to determine the form of the effective hadronic operators from the symmetries of the \BBz-decay operators in a given theory. In all of these cases the estimates of \Mz\ are not as advanced as that for light neutrino exchange and more work is needed.

A number of authors~\cite{Geh07,Dep07,fog09,sim10,fae11} have tried to estimate the number of measurements required to discern the underlying physics. The assumption that is critical to these arguments is that the spread in \Mz\ due to different models reflects the true variation, which is clearly speculative given the uncertainties still remaining in those calculations. With this assumption in mind, the conclusion is that 3 or more experiments with a precision of about 20\% for both experiment and theory are needed to have any hope of discerning the underlying physics. There are other motivations that require multiple \BBz\ experimental results~\cite{Ell06}. The need to prove that the observation is indeed \BBz\ and not an unidentified background is not the least among them. Therefore, a general conclusion from these papers, and those similar to them, indicate that at least 3 (and very likely 4) \BBz\ experiments along with significant theoretical effort are warranted. The need for and utility of several precision experimental results is the critical conclusion.

\section{Toward the Solar Scale}
Atmospheric neutrino oscillation results indicate that at least one neutrino mass eigenstate is about 50 meV or more. In the inverted hierarchy with a small \ml, \mee\ would also be about 50 meV. At this mass the \BBz\ half life would be near $10^{27}$ y. It is this scenario that sets the mass for tonne-scale experiments that will evolve from the projects listed in Section~\ref{sec:expt} and from other R\&D efforts.

In the normal hierarchy, when \ml\ is near 0, \mee\ is 5-10 meV. To reach this level of sensitivity, \BBz\ experiments will require a half-life sensitivity near $10^{29}$ y. This will require 100 tonnes of isotope. The technology for enriching isotopes at this volume is not yet cost effective and research will be required to develop it. Furthermore, energy resolution will be critical to make sure that background from the tail of the \BBt\ spectrum doesn't mask the \BBz\ signal. Techniques to fabricate a very large number of high-resolution crystal detectors in a cost effective way must be developed. Alternatively, improving the resolution of large scintillator detectors will also require an R\&D program.

\begin{figure}[h]
\includegraphics[width=35pc]{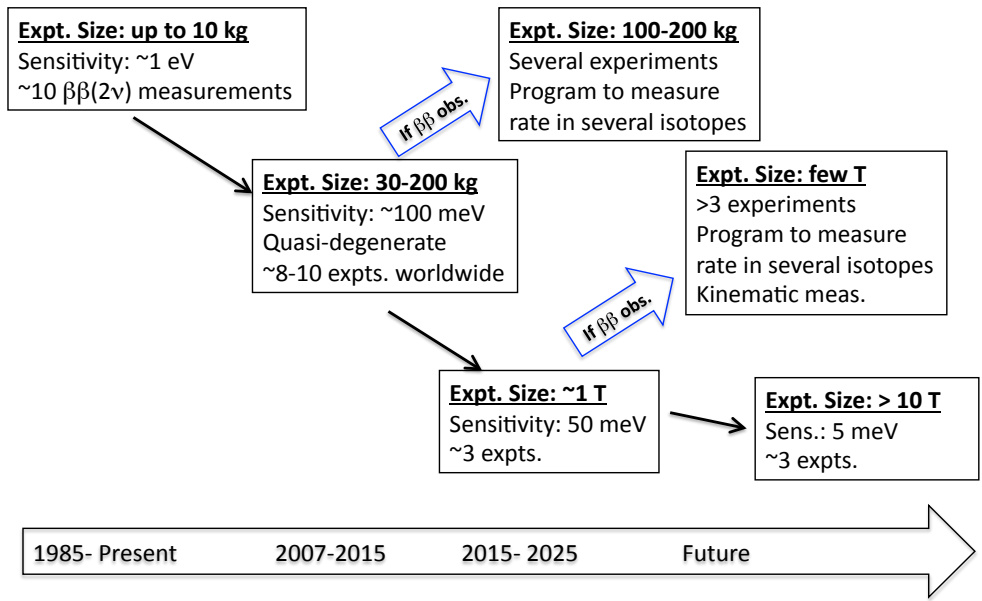}\hspace{2pc}%
\caption{\label{fig:Decision} A possible decision tree based on discovery potential versus limit measurements for the \BB\ program over the coming years. Figure concept derived from discussions with J. F. Wilkerson.}
\end{figure}

As depicted in Fig.~\ref{fig:Decision}, the collection of \BB\ experiments until 2010 used up to $\sim$10 kg of isotope and were successful in measuring \BBt\ in 10 isotopes. In the time period between about 2007 and 2015 experiments based on 30-200 kg will built and operated as discussed in Section~\ref{sec:expt}. With these experiments we anticipate a robust test of the recent claim and either limits or measurements near 100 meV or better. At this point there will be a decision point depending on whether \BBz\ is convincingly detected or not. If it is observed, the program will proceed toward a large number of experiments on this scale. If not, then it will push-on to the tonne-scale and atmospheric mass-scale sensitivity near 50 meV. Another decision point will arise as the results of those experiments. The program will either aspire to even greater sensitivity or a collection of experiments will be developed to better study the decay.

\section{Conclusions}
The worldwide research program in \BB\ is making fast progress due to the great interest in the subject. This interest arises because we now know that neutrinos have mass and that \BBz\ is the only practical way to show that the neutrino is its own anti-particle. That \BBz\ can only exist if neutrinos are massive Majorana particles is a result of virtual exchange of a neutrino. However, to interpret a measured decay rate (or limit) in terms of constraints on neutrino parameters requires an understanding of the matrix elements. \BB\ is a second order weak process that proceeds through many possible intermediate states and therefore is a difficult theoretical challenge. Many measurement efforts are progressing to better understand the nuclear physics involved and better understand these calculations. All told, this is a very exciting time for this field. 

\ack{Acknowledgments}
 I gratefully acknowledge the support of the U.S. Department of Energy through
the LANL/LDRD Program for this work.  I also gratefully acknowledge the support of the U.S. Department of Energy, Office of Nuclear Physics under
Contract No. 2011LANLE9BW. I wish to thank Albert Young for a careful reading of this manuscript and many useful suggestions.

\section{References}
\bibliographystyle{iopart-num}
\bibliography{DoubleBetaDecay.bbl}

\end{document}